\documentclass[runningheads]{llncs}
\usepackage{color,colortbl, latexsym}
\usepackage{graphicx,subfigure}
\usepackage{tabularx}
\usepackage{makecell}
\usepackage{slashbox}

\definecolor{Gray}{gray}{0.7}
\begin{document}
\title{Geo-Enabled Business Process Modeling}
%
%
\author{Behshid Behkamal$^1$ \and Asef Pourmasoumi$^1$  \and Mehdi Akbarian Rastaghi$^1$ \and Mohsen Kahani$^1$ \and Hamid Reza Motahari-Nezhad$^2$ \and Mohammad Allahbakhsh$^{3,1}$ \and Issa Najafi$^4$}
\authorrunning{B. Behkamal et al.}
%
\institute{WTLab, Computer Engineering Department, Ferdowsi University of Mashhad, Mashhad, Iran\\
\email{\{behkamal,asef.pms,kahani,allahbakhsh\}@um.ac.ir}\\
\email{mehdi.akbarian@mail.um.ac.ir}\\
 \and
Ernst \& Young AI Lab, Palo Alto, USA\\
\email{hamid.motahari@ey.com}
 \and
University of Zabol, Zabol, Iran\\
\email{allahbakhsh@uoz.ac.ir}
 \and
National Post Company of Iran, Tehran, Iran\\
\email{issa.najafi@post.ir}}
\maketitle              
\begin{abstract}
Recent advancements in location-aware analytics have created novel opportunities in different domains. In the area of process mining, enriching process models with geolocation helps to gain a better understanding of how the process activities are executed in practice. In this paper, we introduce our idea of geo-enabled process modeling and report on our industrial experience. To this end, we present a real-world case study to describe the importance of considering the location in process mining. Then we discuss the shortcomings of currently available process mining tools and propose our novel approach for modeling geo-enabled processes  focusing on 1) increasing process interpretability through geo-visualization, 2) incorporating location-related metadata into process analysis, and 3) using location-based measures for the assessment of process performance. Finally, we conclude the paper by future research directions.
\end{abstract}
\section{Introduction}
\label{sec:intro}
Process mining is an emerging research area, which helps business analysts and managers gain more insight into their organization’s processes.  Business process mining research has focused on both intra-organizational as well as cross-organizational processes ~\cite{r2}. There are two main types of cross-organizations process mining. In the first, different organizations work with each other on the same instance of a process to complete it, so they act like puzzle pieces; e.g., the process of 'obtaining a building permit'. The latter group includes the processes in which several organizations use a common infrastructure and execute the same process model with some minor customizations; e.g., the process of 'parcel pickup and delivery'. In both groups, the activities of a business process may be performed in different geographic locations. Thus, the location where the process activities are performed can provide helpful information for analyzing process performance.
There are a very limited number of research initiatives that investigate geospatial information for process analysis. The concept of location-dependent task was firstly introduced by Zhu et al.~\cite{r9}. They presented five location-dependent process model patterns to show how location impacts business process modeling. Zhu et al. also proposed a Petri net modeling mechanism, which incorporates location aspects and ways to constrain the execution of activities~\cite{r8,r9}. In another work of the same authors~\cite{r10}, they investigated how location impacts the primary logical relationships in a process control-flow. Also,~\cite{r5} introduced a process mining technique to understand similar movement patterns in an indoor environment.

Our paper reports on our industrial experience in working with business process mining technology while explicitly considering geospatial information. We demonstrate that when business process event logs are enriched with geospatial information of process activities, other peripheral information, which are only obtainable through geolocation can be accessed for performing more advanced business process mining tasks. In the rest of the paper, we first present a motivating scenario in Section~\ref{sec:motiv}. Then, we discuss the shortcomings of available process mining tools and propose our solution in Section~\ref{sec:approach}. Our developed tool suite is presented in Section~\ref{sec:tool}. Finally, we conclude the paper and present future research directions in Section~\ref{sec:conc}.
\section{Motivation}
\label{sec:motiv}
The work reported in this paper is a result of our industrial investigation of business process mining techniques in collaboration with National Post Company of Iran (NPCI). We present a running example based on data provided by our project partner. One of the most recent challenges in NPCI is to identify the best route for parcel delivery across the country so that they can improve their current distribution process. As such, the process that has been reported throughout this paper is a business process named ‘parcel pickup and delivery’. There are four types of activities in this process, each of which is performed at one of the post offices placed in different geographic locations across the country. These activities are: 1) parcel pickup, 2) parcel check-in, when the parcel arrives at a middle station, 3) parcel check-out, when the parcel exits from a middle station, and 4) parcel delivery. All activities are stored in the event logs of the web-based system as presented in Table~\ref{tbl:sample}.
\begin{table*}[h]
  \centering
  \caption{A sample of an event log of our experiment}
  \label{tbl:sample}
\tiny
  \begin{tabular}{ p{1.1cm} p{1.1cm} p{2.1cm} p{1.8cm} p{1.5cm} p{1cm} p{2.7cm}}
   \hline
   Case\_id & Event\_id & Properties& & & \\  \cline{3-7}
            &           & Timestamp& Activity& Resource&City&Location \\  \hline
   1986638&245 &25-05-2017: 11.50 & Parcel pickup      & P.O. 123  & Mashhad  & 37.75888900,45.97833300\\
          &246 &25-05-2017:14.01  & Parcel check-out   & P.O. 123  & Mashhad  & 37.75888900,45.97833300\\
          &247 &26-05-2017: 08.12 & Parcel check in    & P.O. 240  & Tehran   & 37.55527800,45.07250000\\
          &248 &26-05-2017: 15.20 & Parcel check-out   & P.O. 240  & Tehran   & 37.55527800,45.07250000\\
          &249 &27-05-2017: 09.27 & Parcel check in    & P.O. 285  & Shiraz   & 35.84001880,50.93909060\\
          &250 &27-05-2017: 10.02 & Parcel check-out   & P.O. 285  & Shiraz   & 35.84001880,50.93909060\\
          &251 &27-05-2017: 14.38 & Parcel Delivery	   & Postman 12& Shiraz   & 35.12001440,49.93909060\\
   \hline
  \end{tabular}\\
\end{table*}

\begin{figure*}[h]
       \centering
       \includegraphics[scale=0.31]{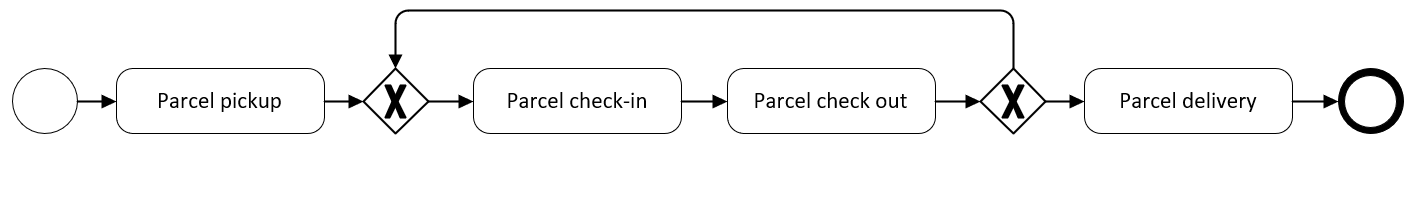}
       \caption{The process model using activity column as 'activity'}
       \label{fig:pmodel}
\end{figure*}

In this log file, the \emph{Case\_id} refers to \emph{Parcel\#} which is a unique number used for tracking, and contains a sequence of activities that are performed to transfer a parcel from source to destination. Besides \emph{Case\_id} and \emph{Event\_id} which are mandatory components of a process instance, there are supplementary information recorded for each case, including \emph{Timestamp, Resource, City}, and\emph{Location}. Notably, the location of each activity was not recorded in the original log file, but we have extracted it from Google Maps using the name of the city. In this experiment, we used a small portion of the event log containing 1,137,643 cases. We first draw a process model using BPMN notation, in which the \emph{Case\_id} is \emph{Parcel\#} and the activities are \emph{parcel pickup}, \emph{parcel check-in}, \emph{parcel check-out} and \emph{parcel delivery}, as shown in Figure~\ref{fig:pmodel}. In this process model, the activity column was selected as \emph{activity}. This provides a view on the flow of the different process steps; however, it cannot present how different process instances are executed.

\begin{figure}[!t]
    \centering
    \subfigure[ProM tool]{
        \includegraphics[scale=0.50]{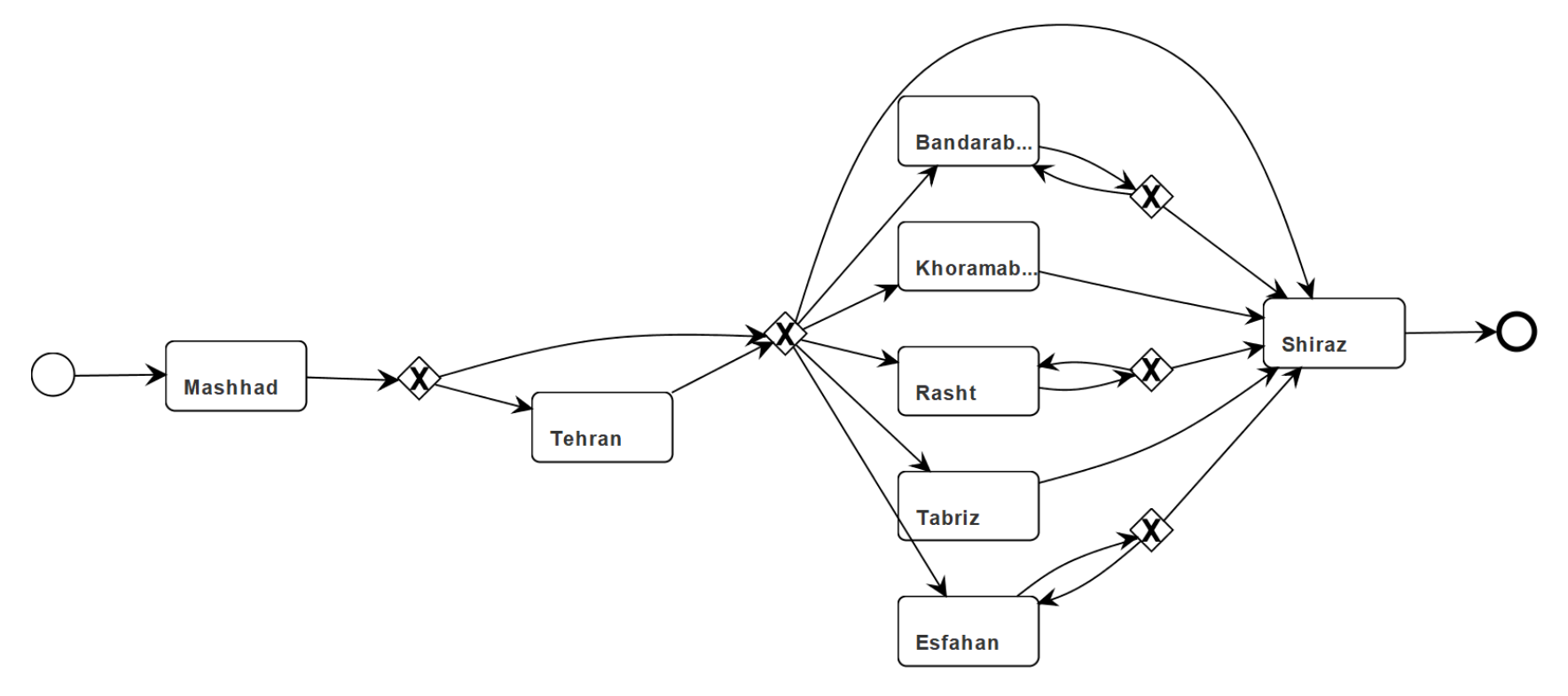}
        \label{fig:prom}
    }
    \subfigure[Disco tool]{
        \includegraphics[scale=0.20]{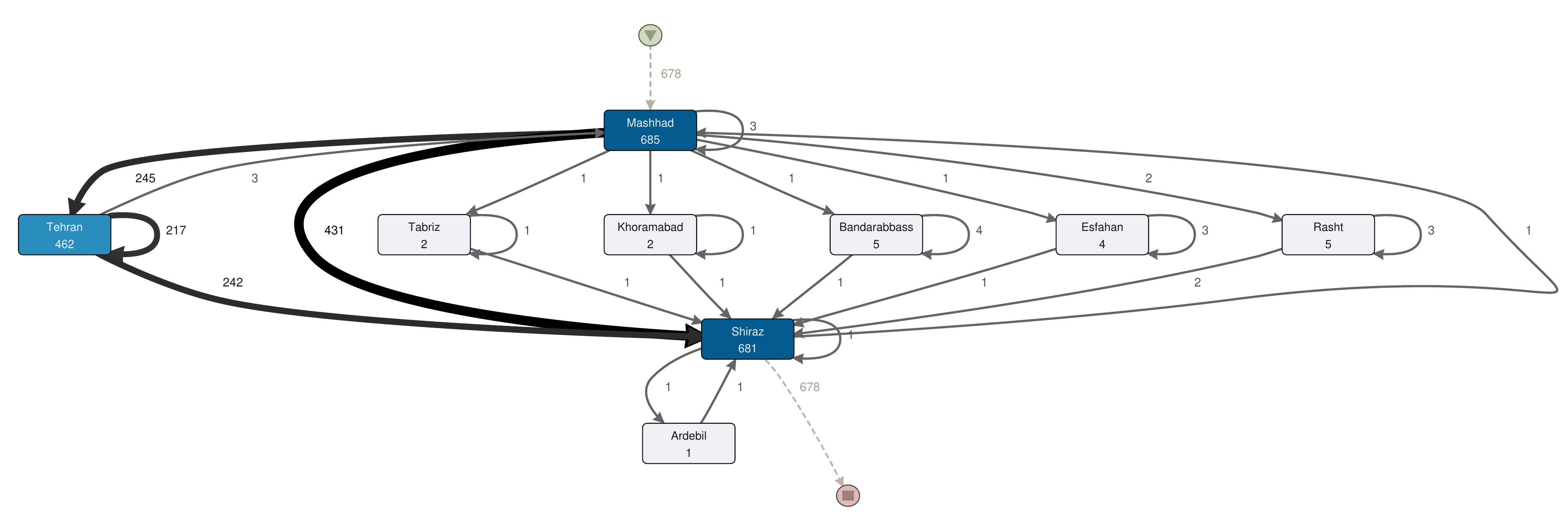}
        \label{fig:pgdisco}
    }
    \caption[]{The process map discovered by ProM and Disco tool using city as 'activity'}
    \label{fig:pgraph}
\end{figure}

Given such a dataset, we can look at the process in a different way by setting another column as \emph{activity}. In this experiment, we consider city as \emph{activity} and automatically discover the process map using two well-known process mining tools, Disco~\cite{r11} and ProM~\cite{r40}. Figure~\ref{fig:pgraph} shows a filtered path of the process map demonstrating only those events which have parcels sent from ‘Mashhad’ to ‘Shiraz’. Although these process models provide valuable insight to analysts and process owners, however, they cannot help with providing a geospatial perspective on the process and answering any geo-related question about the process, such as how can the process of ‘parcel pickup and delivery’ be visually rendered on the map? or what are the paths with the highest degree of parcel traffic across the country? In the next section, we investigate the problems of currently available process mining techniques and introduce our solution for geo-enabled process modeling.

\section{Proposed Approach for Geo-Enabled Process Modeling}
\label{sec:approach}
Based on our practical experience in the analysis of cross-organizational business processes~\cite{r17},  we have realized that currently available techniques do not allow including the geospatial information about process execution into the process models, e.g., the location in which the activities are executed. Here, we are going to briefly present the problems we have practically encountered, and introduce our proposed approach for the geo-enabled visualization of the process.\\

\noindent \textbf{Challenge~1. Complex process models cannot be rendered in an understandable manner.}
A business process can be the visualized by many well-known standards (e.g., Petri Nets~\cite{r30}, Business Process Model and Notation~\cite{r31}, Workflow Nets~\cite{r32}) most of which render process models based on control flow in event logs. There are some methods for designing process models in a way that increases their understandability. For instance, in~\cite{r13}, the authors investigate the effect of layout features of process models on model understandability.\\
\noindent \textbf{Challenge~2. There are valuable location-related metadata that are not considered in process analysis.}
Geospatial information provides access to additional peripheral metadata such as weather and location hierarchy, which are not accessible if geospatial information is not stored in event logs. For example, if we have some information related to the routes such as ground paths, airlines, and railways, or even more detailed information like maximum speed and the number of lanes in a highway, we can effectively mine and analyze a business process log by taking this additional information into account.\\
\noindent \textbf{Challenge~3. Location-based metrics are not currently applied for performance assessment of business processes.}
Typically, three dimensions of performance are identified: time, cost and quality. Time-related information is used to detect deviations and to improve the processes. Also, cost is taken into consideration for performing process analysis~\cite{r18}. However, addition of geospatial information to event logs makes it possible to consider new metrics for the assessment and enhancement of business processes. For example, in Figure~\ref{fig:pgraph}, if there are two routes between 'Mashhad' and 'Shiraz' with different path lengths, current tools cannot explain why it takes more time to deliver the parcel via the longer path from 'Mashhad' and 'Shiraz'. In other words, it is not possible to use the geospatial information as a metric to assess business process performance.

\begin{figure*}[h]
       \centering
       \includegraphics[scale=0.7]{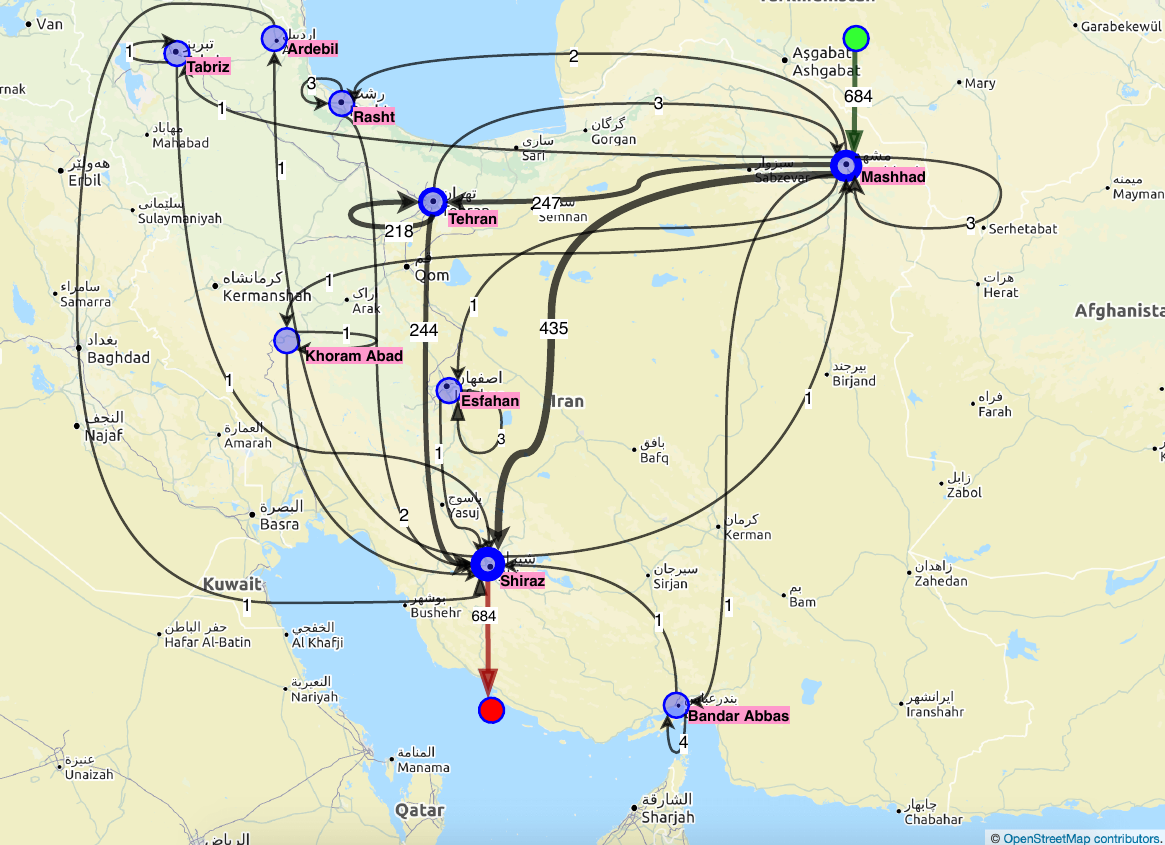}
       \caption{The geo-enabled business process model}
       \label{fig:geoviz}
\end{figure*}

To address these problems, we propose a novel modeling approach for geo-enabled processes.  We define a geo-enabled process model consisting of four perspectives: 1) process perspective focusing on the control flow of geo-located activities, 2) organizational perspective, which depicts the organizational structure to represent geo-visiulized inter-organizational activities, 3) case perspective, showing the characteristics, including geospatial features, of a process instance, and 4) event perspective, which represents the timing, the order, and the location of the events. In other words, for geo-enabled processes, we add the geospatial perspective of business processes focusing on the geographic location of process activities. In a geo-enabled process model, the nodes of the model represent real-world geographic locations and the transitions between nodes represent the activities taken place between different nodes. Given different scales of locations (from coarse to fine-grained), the process model may be defined hierarchically, with regard to the location, and include hierarchical locations in multiple scales. We may overlay the geo-enabled process model with additional layers on the map to represent location-related data from external data sources to allow for better analysis of the process. We have implemented these abstractions and illustrate the use of such additional layers in the next section. Figure~\ref{fig:geoviz} illustrates one realization of a geo-enabled business process model on a map.

\section{Geo-enabled Process Modeling in Practice }
\label{sec:tool}
To bridge the gap between the process execution in the real world and the process model discovered from event, we have developed a tool suite, called MEEM, using the ‘Open Street Map API’ platform with an open-content license. MEEM stands for ‘Meeting of Events and Evidence on the Map’, which integrates two different aspects of data, i.e., ‘\emph{Events}’ and ‘\emph{Evidence}’. On the ‘\emph{Events}’ perspective, the information related to the process activities is extracted from the geo-tagged event log presented in Table~\ref{tbl:sample}. Two layers are implemented in the ‘\emph{Events}’ perspective to show the overview of parcel delivery across the country in both static and dynamic fashions. We have also, implemented the '\emph{Evidence}’ perspective in which valuable location-related data available in external data sources are collected and visualized in different layers, such as main roads, airlines and railways.  The online demonstration of the tool, repositories, and other dissemination materials are publicly available at \emph{ https://github.com/makbn/meem}. Here, we present the advantages of geo-enabled process modeling and address the challenges mentioned in the previous section.

\begin{figure}[!t]
    \centering
    \subfigure[The discovered process map from event log using Disco.]{
        \includegraphics[scale=0.35]{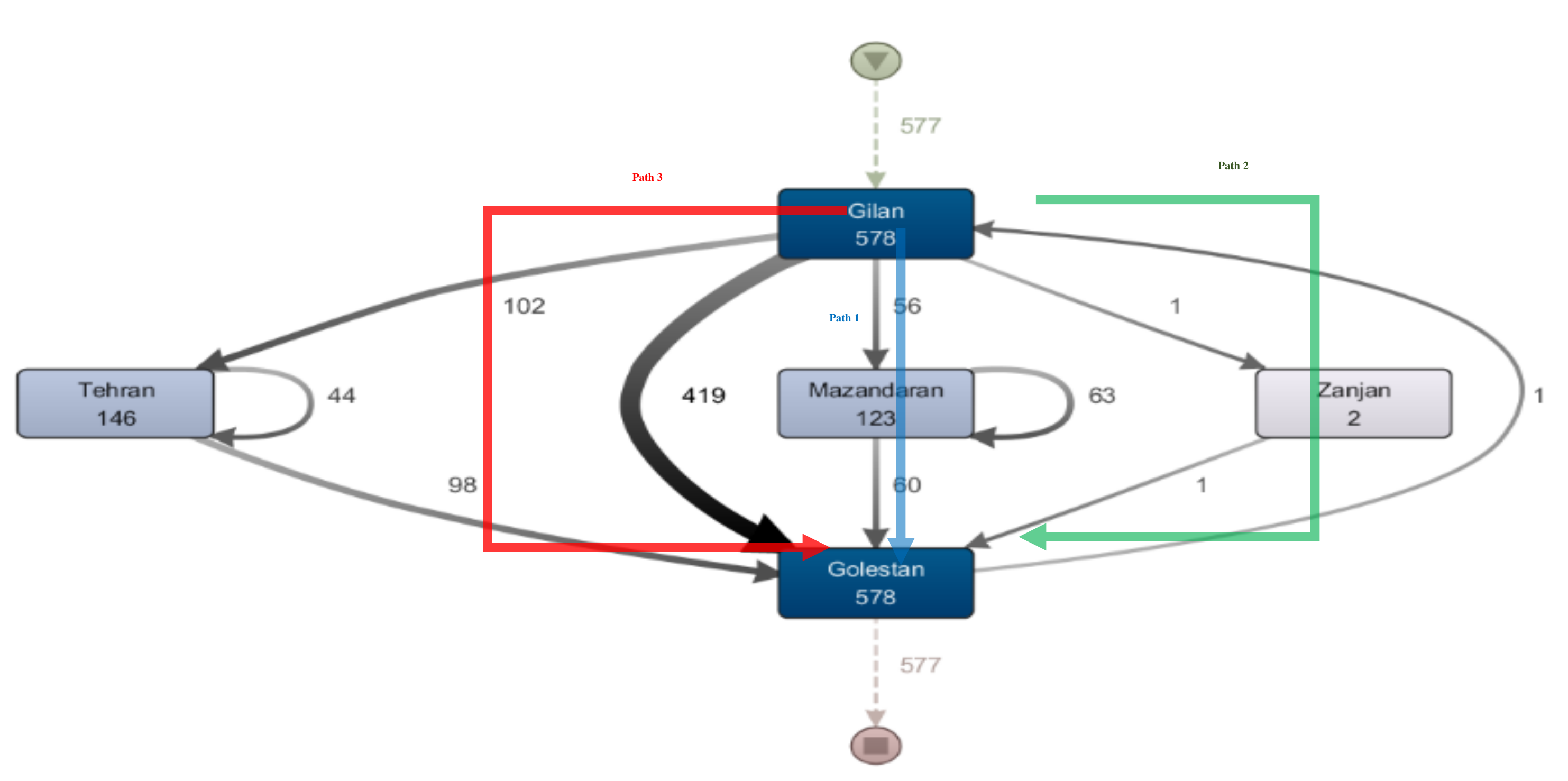}
        \label{fig:disco}
    }
    \subfigure[ The discovered process map from geo-tagged event log using our tool.]{
        \includegraphics[scale=0.3]{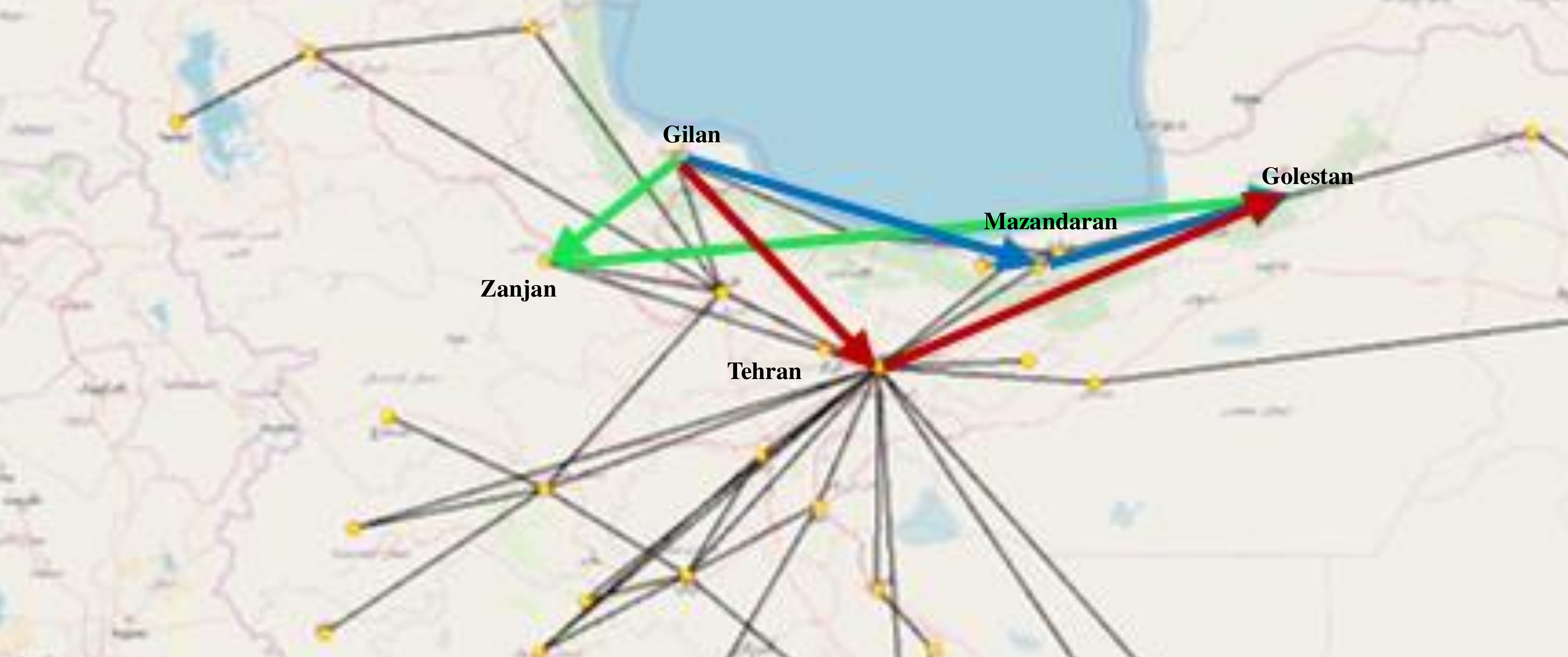}
        \label{fig:meem}
    }
    \caption[]{The filtered paths from `Gilan' to `Golestan'}
    \label{fig:scn12}
\end{figure}

\noindent \textbf{S1. Increasing process interpretability.}
One of the challenges of current process mining tools is to understandably draw the workflow of a process. Given geospatial information, we are able to address this issue and demonstrate how a process model would be more understandable through geovisualization. To this end, we first draw the workflow of the parcel delivery process without considering geo-tags using Disco~\cite{r11}. Then, we depict the same process model on the map and explain its main advantages compared to the former. To visualize the workflow of the process, we selected two Iranian provinces, namely ‘Gilan’ and ‘Golestan’, as the source and destination, respectively. Then, we filtered for only those events which have parcels sent from ‘Gilan’ to ‘Golestan’ using ‘Endpoints Filter’ option of Disco shown in Figure~\ref{fig:disco}. We then depict the process graph discovered from geo-tagged event logs using our tool which is shown in Figure~\ref{fig:meem}. It is clear that when the process graph is visualized on a map, it is more understandable as the activities (nodes of the graph) are placed on real-world geographic locations.

\noindent \textbf{S2. Incorporating location-related properties into process models.}
We believe that process mining will benefit from location hierarchy. For example, we are able to filter the events at different levels of post offices, cities, and countries. So, we have used location hierarchy as metadata to address the second challenge.

\noindent \textbf{S3. Using location-based measures for the assessment of process performance.}
Considering geospatial information will enable process analysts to improve the efficiency of process execution. For example, the distance between the locations of two consecutive activities can be used as a measure for the assessment of process performance along with the execution time. As shown in Figure~\ref{fig:disco}., there are three routes for parcel delivery from ‘Gilan’ to ‘Golestan’, which are drawn in different colors. Based on the thickness of the arrows (that shows the number of parcels which have traveled through that path) we can see that most of the parcels are transferred via Path 1. However, by this representation of the process graph, we cannot detect which path is optimal since the distances between nodes are not presented in this model. When the process graph is drawn using geo-tagged data, it is possible to place the activities on the exact location they are executed. As such, not only the execution time of a process instance is considered for the assessment of process performance, but also the distance between activity locations can be applied as a complementary metric to better analyze the process execution. Thus, we can simply infer that among all three existing paths from ‘Gilan’ to ‘Golestan’, Path 1, which is drawn in blue, is the optimal path in terms of distance between source and destination.

\section{Conclusion and Future Research Directions}
\label{sec:conc}
The goal of this paper is to contribute to process mining by introducing ‘geo-enabled process modeling’. We have first motivated our work by presenting a real world scenario and enumerating some challenges faced by available tools. Then, we have presented our developed tool and demonstrated how process mining can benefit from adding the geospatial dimension in practice. There are two promising research directions: 1) visualizing the formal process model on the map. Here, we have just confined to drawing a simple representation of the process graph on the map. To geovisualize the formal process model, the existing plug-ins such as Alpha Miner and the Inductive Visual Miner will be used to create new plugins; 2) Proposing a novel location-based metric. Most of the process mining techniques calculate performance based on execution time. Adding the new location dimensions to event logs, we can better analyze the performance of business processes using a distance-based metric.

\bibliographystyle{unsrt}
\bibliography{refs}

\end{document}